\documentstyle[prl,aps,epsf,multicol]{revtex}
\draft

\title{Magnetization reversal triggered by spin injection in 
magnetic nanowires}
\author{J.-E. Wegrowe \cite{email}, D. Kelly, Ph. Guittienne, J-Ph. Ansermet}
\address{Institut de 
Physique Exp\'{e}rimentale, Ecole Polytechnique F\'{e}d\'{e}rale de 
Lausanne, CH-1015 Lausanne, Switzerland}

\begin{document}

%\pacs{PACS numbers: 75.40. Gd, 75.70. Pa, 66.20 \hfill}

\maketitle

\begin{abstract}
It is shown that a pulsed current driven through Ni nanowires provokes 
an irreversible magnetization reversal at a field distant from the 
spontaneous switching field $H_{sw}$ by $\Delta H$ of as much as 40 \% 
of $H_{sw}$.  The state of the magnetization is assessed by 
magnetoresistive measurements carried out on single, isolated 
nanowires.  The reversible part of the magnetization follows that of a 
uniform rotation.  The switching occurs between the two states 
accessible otherwise by normal field ramping.  $\Delta H$ is studied 
as a function of the angle between the applied field and the wire, and 
also of the direction of the pulsed current.  The results are 
interpreted in terms of spin-flip transfer from the 
spin-polarized current to the magnetization, while the switching is 
approximated by a curling reversal mode.
\end{abstract}

\section{Introduction}
Is it possible to trigger magnetization reversal without the need of a 
magnetic field?  Some recent studies in spin electronics suggest that 
such an effect may now be evidenced and controlled.

Spin dependent scattering studies emerged with the first realizations 
of magnetic nanostructures, and gave rise to the discovery of spin 
injection \cite{Johnson}, giant magnetoresistance (GMR) \cite{GMR} and 
tunneling magnetoresistance (TMR) \cite{TMR}.  Thanks to the 
nanoscopic scale of these artificial magnetic systems, the properties 
of the spins carried by the conduction electrons can be exploited.  In 
particular, the electric response is determined by the magnetization 
states \cite{Prinz}.  This paper addresses the reverse effect, namely 
the ability of controlling the magnetization states with the spins of 
the conduction electrons.  Some years ego, Berger predicted the 
existence of phenomena due to the action of spin polarized conduction 
electrons on domain walls or on spin waves \cite{Berger} in magnetic 
thin films.  Recently, Slonczewski predicted the rotation of the 
magnetization due to polarized currents in multilayered systems 
\cite{Sloncz}, and Bazaliy et al.  derived from microscopic 
considerations a generalized Landau-Lifshitz-Gilbert (LLG) equation 
\cite{Bazaliy}. A macroscopic derivation of the generalized LLG 
equation based 
on thermokinetics principles is proposed in reference \cite{16}. 
 From an experimental point of view, Freitas and 
Berger and Hung and Berger \cite{ExpeBerger1} showed the action of a 
high current density on domain walls in thin films.  Recent 
experiments on nanostructured samples bring new evidence for an 
interpretation in terms of the action of the spin of the conduction 
electrons.  Tsoi et al.  \cite{Tsoi} showed the effect of a high 
current density on spin wave generation in Co/Cu multilayers and Myers 
et al.  reported an effect of current induced switching in magnetic 
multilayer devices \cite{science}.  The effect was evidenced in these 
two experiments by the observation of peaks and hysteretic behavior of 
the value of the differential resistance dV/dI. We report here the 
direct observation of current-induced magnetization reversal, using 
for this study well defined magnetization states whereas earlier work 
involved domain wall configurations \cite{EPL}.  

\section{Measurements}
The experiments are performed on magnetic nanowires at room 
temperature.  This study focuses on the irreversible part of the 
hysteresis where the magnetization defines a two state system.  The 
irreversible transition has been largely studied experimentally and 
theoretically in such microstructures, thanks to a high aspect ratio 
favoring magnetic single domain configuration 
\cite{Aharoni}\cite{M(H)}\cite{PRL}.

The samples were produced with the method of electrodeposition in 
track etched membrane templates.  Gold layers were sputtered on both 
sides of a nanoporous membrane and the electrodeposition of Ni were 
performed.  The wires were about 70 nm in diameter and 6000 nm in 
length.  A single contact was obtained by monitoring the potential 
between both sides of the membrane during the electrodeposition 
\cite{IEEE}.  The wires were characterized by TEM and X-ray 
diffraction \cite{Meier}.

\begin{figure}
\epsfbox{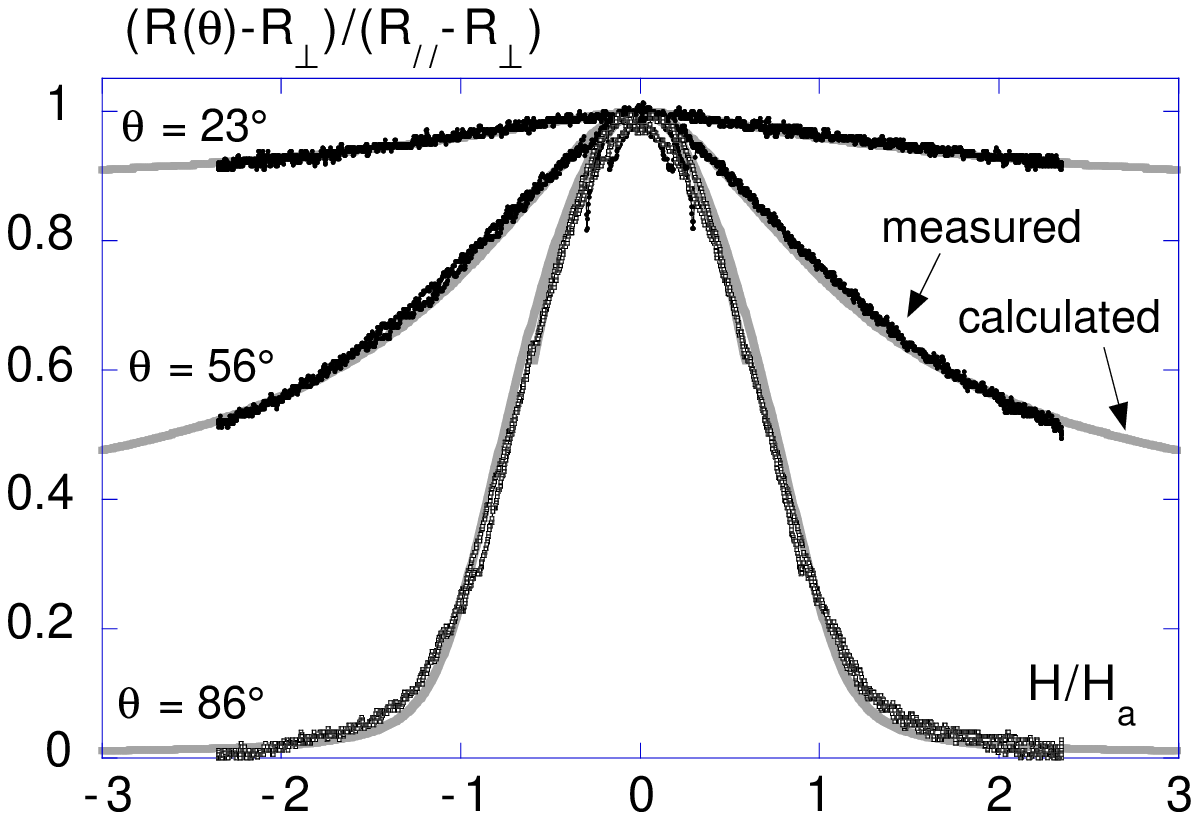}
\vspace{0.5cm} \caption{Anisotropic magnetoresistance 
measured with a current of about $10^{4}\,A/cm^2$, at different 
angle $\theta $ of the applied field.  Grey lines are calculated 
curves assuming uniform magnetization.  $R_{\|}$ is the resistance 
with the magnetization parallel to the current, $R_{\bot}$ is the 
resistance with the magnetization perpendicular to the current.  
$\mu_{0} H_{a}\,=\,$ 300 mT.}
\label{hyst}
\end{figure}

The consecutive magnetization states of a single Ni nanowire are 
measured through magnetoresistance curves (Fig.~\ref{hyst}) by ramping 
the field for different values of the angle $\theta$ of the applied 
field with respect to the wire axis.  The magnetoresistive curves are 
measured with a current density of $10^4$ A/cm$^2$ at a magnetic field 
sweep rate of 1 mT/s.  The anisotropy field calculated from the 
demagnetizing factors of the wire is about $\mu_{0}H_{a}\,=\,300\,$mT. 
The irreversible part of the hysteresis loop consists simply of two 
symmetric jumps of the magnetization, one for the decreasing field and 
one for the increasing field.  Magnetic characterizations of 
electrodeposited Ni nanowires through AMR measurements were reported 
in details in \cite{PRL}, \cite{Yvan}, \cite{Ferre}.  In the case of 
uniform magnetization, a simple quadratic relation between the 
magnetoresistance hysteresis loop $R(\vec{H})$ and the magnetic 
hysteresis loop $M_{z}(\vec{H})$ holds \cite{AMR} $ R(\vec{H}) = 
R_{0}+ \Delta R_{max}.\,(M_{z}(\vec{H})/M_{s})^{2}$, where $\Delta 
R_{max} $ is the difference between the resistance with magnetization 
parallel and magnetization perpendicular to the current and $M_{s}$ is 
the saturation magnetization.  The calculated curves $R(\vec{H})$ 
\cite{Yvan}, plotted with the experimental data in (Fig.~\ref{hyst}), 
show that the profile of the reversible part of the hysteresis follows 
that of the uniform rotation.  Hence, magnetic inhomogeneities, if any 
are present, constitute a few percent of the magnetization at most.

The purpose of this work is to know what happens when a pulsed current 
is injected close to the irreversible part of the hysteresis loop.  
The following protocol is used: the external field is sweept slowly up 
to a given value of the applied field $H$ close to the regular 
switching field $H_{sw}(\theta)$ at which the irreversible jump of the 
magnetization occurs spontaneously.  At this fixed field, a pulse of 
an amplitude of about $I_{e}\, \approx \, \pm 1$ mA (i.e. about $2.6 
\cdot 10^{7}\,$A/cm$^2$) and $0.5 \mu s$ duration is injected in the 
wire.  The effect of the pulse is to provoke the magnetization 
reversal from the stable state defined by the applied field 
$H$ to the next stable state located on the symmetric 
half-hysteresis loop, i.e. on the branch corresponding to the 
succession of magnetization states obtained with a decreasing applied 
field.  This protocol is then repeated for other values of $H$, 
until a maximum distance $\Delta H_{max}\, = \,|H - H_{sw}|$ is 
reached, beyond which the pulsed current does not affect the 
magnetoresistive loop any longer.  A detailed view of the irreversible 
part of the loop, measured at $ \theta = 65^o$ is shown in 
Fig.~\ref{Ipulse}, (continuous line).  The dashed lines show the 
magnetoresistance with injection of pulsed current at various value of $ 
H$.  This protocol is repeated for different angles of the applied 
field (Fig.~\ref{DHswI}), and for the two directions of 
the pulsed current.

A clear asymmetry is observed between positive and negative currents.  
There is no noticeable asymmetry with respect to the direction of the 
external field, that is, for a given sense of the current, the same 
effect is seen for both branches of the hysteresis loop (the half 
hysteresis loop with decreasing field and the half hysteresis loop 
with increasing field).  We must attribute \cite{sym} this asymmetry 
of the effect with the current direction to differences in the ends of 
the nanowires.  The ends are indeed morphologically different.  The 
interface at the top of the membrane (where the growth ends) contains 
the micro-contact.  It forms a Ni mushroom-like growth \cite{Meier} 
spreading over the 80nm gold layer.  The interface at the bottom of 
the membrane is a 150nm thick gold layer onto which the nucleation 
occurs during the first stages of the Ni electrodeposition.

\section{Origin of current-induced magnetic switching}
What is the mechanism responsible for the switching of the 
magnetization : the field induced by the current, Joule heating, or 
the effect of spin-polarized currents?

\begin{figure}
$$
\epsfbox{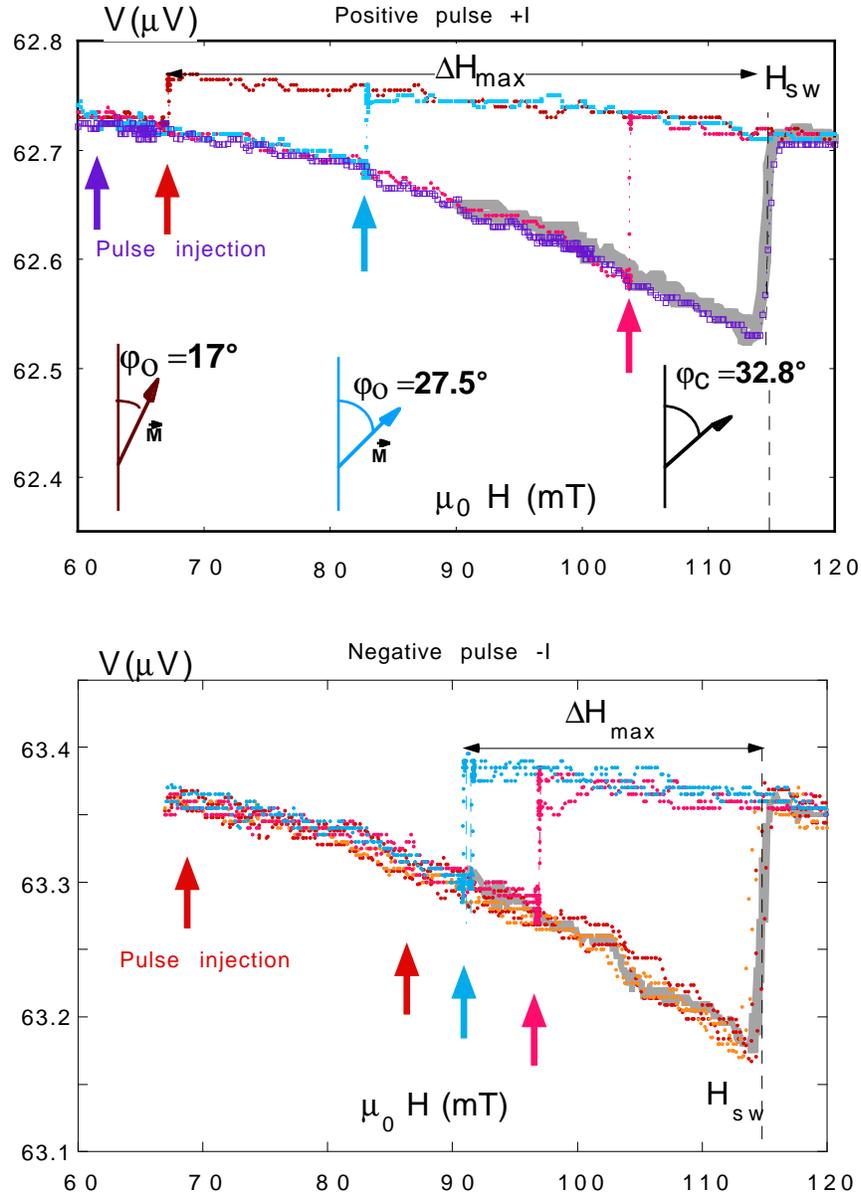}
$$
\caption{Irreversible part 
of the magnetoresistance measured with $10^{4} A/cm^{2}$ reading 
current (grey line) at $65^o$ for the two directions of the pulsed 
current.  Pulsed current of $2.6 \cdot 10^{7} A/cm^{2}$ amplitude and 
$ 0.5 \mu sec$ duration are injected at various applied field $ 
H$, inducing a jump of the magnetization.  Above $\Delta H_{max}$, 
there is no more effect of the pulsed current, and the 
magnetoresistance follows the regular curve without pulse.  (a) 
Current +I.  Three configurations of the magnetization, defined by the 
angle $\varphi_{0}$, are sketched, up to the critical angle 
$\varphi_{c}$ where the magnetization is no more stable. (b) Opposite current.}
\label{Ipulse}
\end{figure}

\begin{figure}
$$
\epsfbox{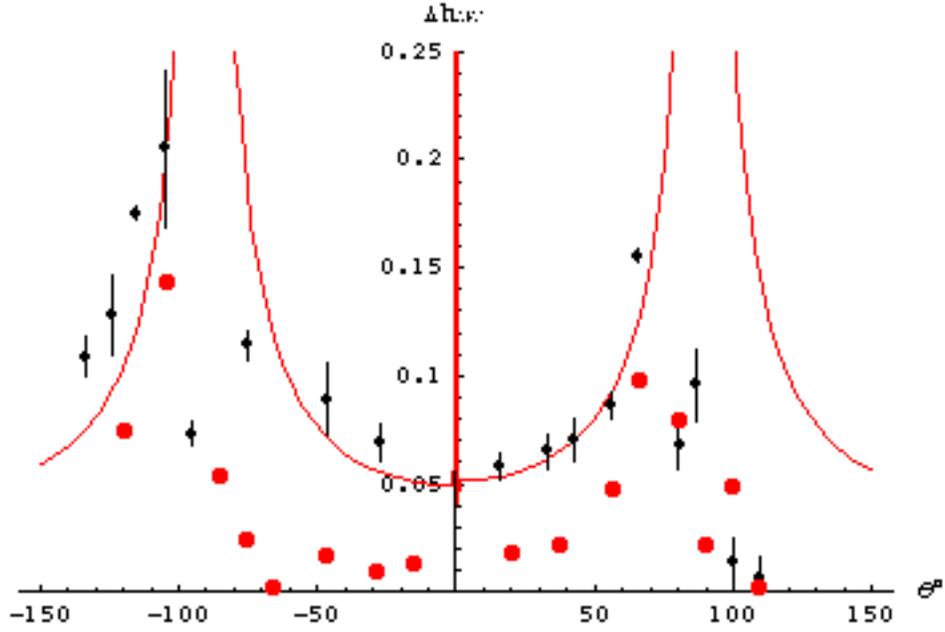}
$$
\caption{Angular dependence of the parameter $\Delta h\,=\, \Delta 
 H_{max}(\theta)/H_{a}$ for pulsed current of $\pm 1mA$ (about $2.6 
 \cdot 10^{7}\,A/cm^2$).  Small dots with error bars: positive 
 current.  Grey points: negative current.  The curve is given by the 
 Eq.~(\ref{result}) of the text, with $ c\,=\,$500 A$^{-1}$.}
\label{DHswI}
\end{figure}

The contribution of Joule heating to $\Delta H_{max}$, through 
enhancement of the thermal activation, can be estimated to be about 3 
mT variation on the switching field (due to a temperature rise of 
about 10K).  But a definitive argument against an interpretation in 
terms of thermal activation due to Joule heating is the asymmetry 
observed for the two directions of the current (see Fig.~\ref{Ipulse} 
and Fig.~\ref{DHswI}): the thermal power dissipated in the wire does 
not depend on the direction of the current whatever the magnetic 
configurations.

The evaluation of the effect of the circumferential field induced by 
the pulsed current on the magnetization reversal can be estimated here 
because the magnetization is uniform before the irreversible jump.  The 
maximum induced field $H_{circ}\,=\,\frac{I}{2 \pi r}\,$A/m is about 5 
mT for 80 nm diameter, i.e. about one order of magnitude smaller than 
the measured effect $\Delta H_{max}$.  Beside this evaluation, the 
experimental argument which invalidates the hypothesis of the induced 
field comes from the study of $\Delta H_{max}$ as a function of the 
wire diameter in the range 120 nm downto 18nm, for a set of 
samples with curling-like reversal modes.  The maximum of $\Delta 
H_{max}(\theta)$, of the order of 50 mT for 80 nm diameter, decreases 
down to about 40 mT for diameters below 40nm at constant current I=1mA 
(curve not shown in this letter), in contradiction with the behaviour 
expected for the induced field.  Note also that the gradient of such 
induced field is two orders of magnitude less than the gradient of 
field in a typical domain wall in nickel.

The most important feature of the measurements shown in 
Fig.~\ref{Ipulse} is that for all applied field $H$ the final 
states after the current pulse are located on the same hysteresis curve 
(corresponding to uniform configurations as shown in Fig.~\ref{hyst}) 
as in the case of the switch without current pulse.  In contrast to 
our previous study of non-uniform nagnetization states \cite{EPL}, the path followed 
by the magnetization during the current pulse may then be described in 
three steps.  First, the magnetization rotates uniformally from the 
equilibrium state (e.g. the position $\varphi_{0}= 17^o$ in 
Fig.~\ref{Ipulse}) to the critical state (defined by the position 
$\varphi_{c}= 32.8^o$ in Fig.~\ref{Ipulse}) where the magnetization is 
no longer stable; second, 
irreversible magnetization reversal (curling mode); finally, the magnetization 
rotates down to its equilibrium value defined by the external field, 
following the back half hysteresis loop.

These observations introduce the question as to how to describe the 
effect of the spin of the conduction electrons.  Due to spin-flip 
scattering, the spin polarization of the current just before entering 
a magnetic layer is not equal to the spin polarization of the current 
inside it.  There is clearly no conservation of spin polarization of 
the current \cite{Sloncz}, \cite {Bazaliy}.  The generalized 
Landau-Lifshitz equation with the additional term describing the 
change in the magnetization due to the spin-flip scattering inside the 
magnet can be 
written as, \cite{16},

\begin{eqnarray}
\frac{d\vec{M}}{dt}\,&\approx &\,-\,g'M_{s}\left(\vec{M}_{0} \times 
\frac{dV}{d\vec{M}_{0}}\right) \nonumber\\
& &-h' \left(\vec{M}_{0} \times \frac{dV}{d\vec{M}_{0}}\right) \times 
\vec{M}_{0}\,+\,p \frac{g\mu_{B}\beta I_{e} \vec{e}_{p}}{eL}
\label{GLL}
\end{eqnarray} 

 where the first, second and third term in the right hand side are 
 respectively, the precession term (or transverse relaxation), the 
 longitudinal relaxation term, and the maximum spin injection due to 
 spin polarized conduction electrons.  $M_{0}$ is the magnetization of 
 the wire without current, L is the length of the wire, e is the 
 electric charge of the electron, $ \beta $ is the bulk conductivity 
 asymmetry \cite{17}, p is a geometric factor, and the unit vector $ 
 \vec{e}_{p}$ gives the direction of the spin polarization of the 
 incident current.  The phenomenological parameters h' and g' 
 \cite{Coffey} are linked to the gyromagnetic ratio $ \gamma $ and the 
 Gilbert damping coefficient $ \alpha $ by the relation 
 $h'\,=\,\frac{\gamma \alpha}{(1+\alpha^2)M_{s}}$ and 
 $g'\,=\,\frac{\gamma}{(1+\alpha^2)M_{s}}$.
 
\begin{figure}
\epsfbox{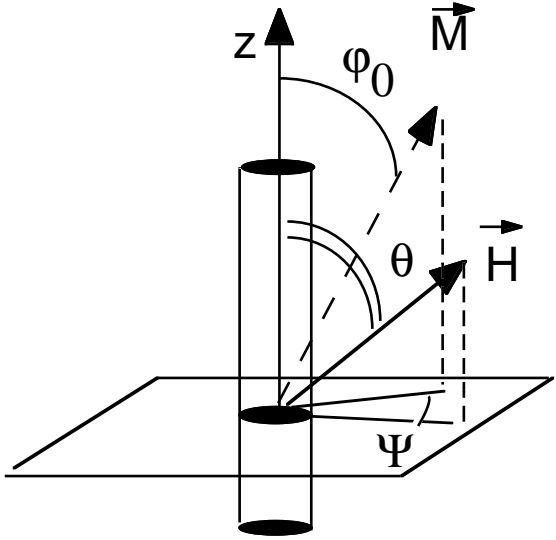}
\caption{Uniform magnetization and the 
magnetic field in the case of uniaxial anisotropy: definition of the 
angles.}
\label{angles}
\end{figure}

In order to estimate the measured parameter $\Delta H_{max}$ , 
we assume a uniform magnetization $\vec{M}\,\approx \, M_{s}\vec{u}$
which defines the unit vector $\vec{u}$ at an angle $\varphi$ of the 
wire axis (see Fig.~\ref{angles}).  The Gibbs energy density can then 
be written in the following form :
\begin{eqnarray}
V(\varphi,\, \Psi)\, = \,
KS & &\,(-cos^2 \varphi \,- \, 2h\, (cos(\theta)cos(\varphi) 
\nonumber\\
& &+\,sin(\theta)sin(\varphi)cos(\Psi)))
\label{potential}
\end{eqnarray}

where $h \,= \,\frac{H}{H_{a}}$ is the applied magnetic field 
normalized to the anisotropy field, 
$K\,=\,\frac{\mu_{0}}{2}\,H_{a}M_{s}$ is the shape anisotropy, and S 
the section of the wire and $\Psi $ is the out of plane coordinate of 
the vector $\vec{u}$ .  The cylindrical geometry of the wires implies 
that $\Psi $=0.

In our experimental protocol, the angle reached by the magnetization 
during the current pulse is the critical angle $\varphi_{c}$ at which 
the irreversible jump of the magnetization occurs.  The maximum 
distance $\Delta h\,= h_{sw}\,-\,h$ where the jump of the 
magnetization can still be observed corresponds then to the variation 
of the angle $\Delta \varphi = \varphi_{c}-\varphi_{0} $ needed to 
shift the magnetization up to the unstable state.  The angle 
$\varphi_{0}$ is given by the equilibrium condition $\left( 
\frac{dV}{d\varphi}\right)_{\varphi_{0}}\,=\,0$.  For steady states 
and neglecting the precessional term (low frequency measurements and 
high damping limit) Eq.~(\ref{GLL}) and Eq.~(\ref{potential}) lead to 
\cite{16}:

\begin{equation}
\Delta h\,=\, h_{sw}(\theta )\,-\, 
\frac{cI_{e}\,(\vec{e}_{p}.\vec{v})
\,-\,sin(2 \varphi_{c})}{2 \, sin(\varphi_{c} \,-\, \theta)}
\label{result}
\end{equation} 

where $\vec{v}$ is the polar vector perpendicular to $\vec{u}$ in the 
$\Psi $=0 plane.  The parameter c is defined by the relation

\begin{equation}
c\,=\,\frac{p \beta \, \hbar}{eKv_{a} \alpha}
\label{cparameter}
\end{equation}

where the activation volume $v_{a}$ of magnetization $M_{s}$ was 
estimated to be
$v_{a}\,\approx \,10^{-22} \,m^3$, $K\,\approx\,10^5 J/ m^3$ 
\cite{PRL}, and
$\, p\, \beta \, \approx \, 0.3 $ \cite{17}, 
 $\alpha \,\approx\,0.07$ \cite{18}. We obtain $c\,\approx\,400 
 A^{-1}$.

All parameters in Eq.~(\ref{result}) are known if the magnetization 
reversal mode, which describes the irreversible jump, is known.  In 
some few theoretical models of magnetization reversal \cite{Aharoni}, 
the functions $H_{sw}(\theta)$ and $\varphi_{c}(\theta)$ are 
analytical.  In the framework of the present empirical approach, the 
experimental data are analysed using the relation deduced from a 
curling 
reversal mode \cite{M(H)} \cite{PRL} :

\begin{equation}
h_{sw}(\theta)\, =\,\frac{a(a+1)}{\sqrt{a^{2}+(2a+1)cos^{2}(\theta)}}
\label{curling}
\end{equation}

\begin{figure}
\epsfbox{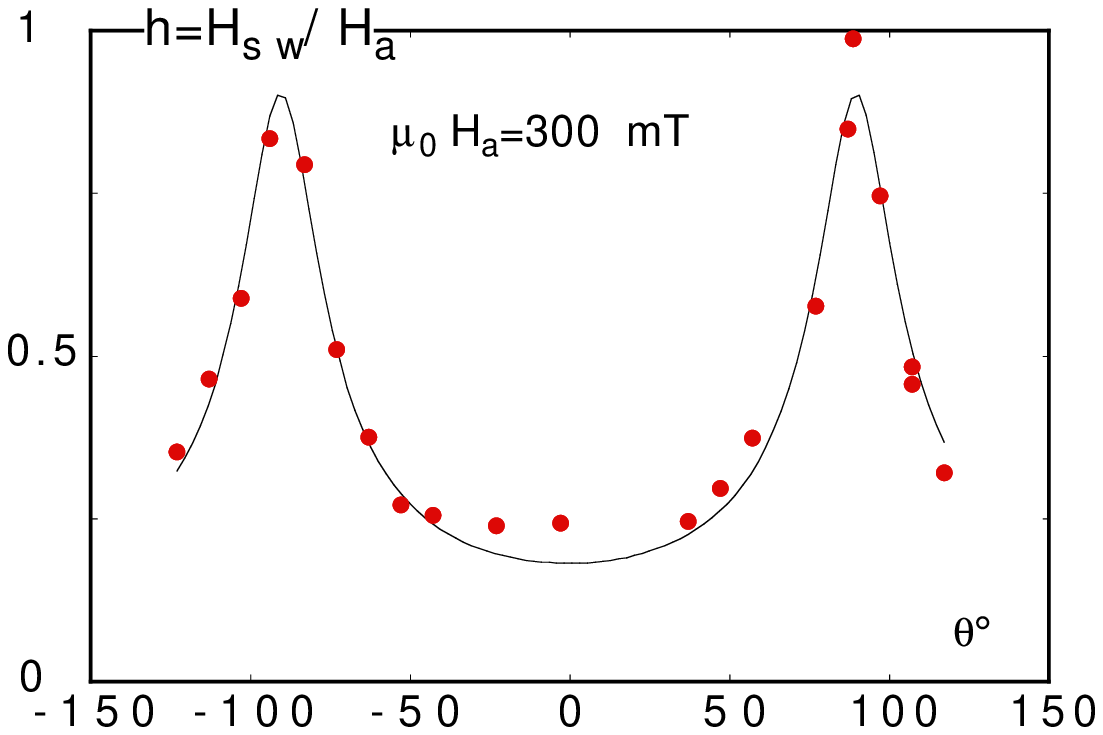}
\vspace{+0.5cm}
 \caption{Circles: measured position of the switching 
field $H_{sw}$ for different angle of the applied field. Line: One 
parameter fit with the curling formula Eq.~(\ref{curling}).}
\label{Hsw}
\end{figure}

The single adjustable parameter $a \,=\, -k \,(R_{0}/r)^2$ is defined 
by the geometrical parameter k \cite{Aharoni}, by the exchange length 
$R_{0}\, =\, 20 \,nm$ ,\cite{EPL} and by the radius of the wire r.  
The fit to the experimental points for $H_{sw}(\theta)$ ( 
Fig.~\ref{Hsw})
 yield $a\, =\, -0.15$ (which corresponds to r of about 60 nm).  
 The relation between the angle of the applied field $\theta$ and the 
 angle of the magnetization $\varphi^{c}$ is:

\begin{equation}
tan(\theta)\,=\,\frac{a+1}{a}tan(\varphi_{c})
\label{curling2}
\end{equation}

We assume for reasons of symmetry (which define the demagnetizing 
field) that $\vec{e}_{p}$ is in the direction of the wire axis 
:$\vec{e}_{p}.\vec{v}\,=\, sin(\varphi_{0})$.  The curve $\Delta h$, 
evaluated from Eq.~(\ref{result}) by numerical resolution is plotted 
in Fig.~\ref{DHswI}, together with the experimental data.  The best 
value of the adjustable parameter c is about $c\,=$ 500 A$^{-1}$ , 
which is in accordance with the rough evaluation of 
(\ref{cparameter}).  The divergence at $90^o$ is due to the numerical 
resolution of Eq.~(\ref{result}) (in which numerator and denominator 
tend to zero).  The same fit to the data obtained with the opposite 
current gives a parameter $c\,=$ 200 A$^{-1}$ (curve not plotted in 
Fig.~\ref{DHswI}).  The comparison between the data and the model in 
Fig.~\ref{DHswI} shows that the phenomenon occurs as if the 
magnetization reversal were provoked by the spin transfer from the 
incident current into the ferromagnet, with a spin polarization in the 
direction of the wire axis.  The observed asymmetry in the sense of 
the current means that the polarization of the incident current is 
different if the current flow is oriented from bottom up or from top 
to bottom.

\section{Conclusion}
The effect of pulsed current on the irreversible magnetization 
reversal was measured.  The amplitude of the effect is more than a 40 
\% variation of the switching field.  In other words, it corresponds 
to a change of the orientation of the magnetization more than $\Delta 
\varphi = 12^o$ for a current of about $2.6 \cdot 10^{7}\,$A/cm$^2$.  
The effect is interpreted in terms of spin flip scattering.  The 
origin of the spin polarization of the current could not be 
unequivocally evidenced, but the assymmetry with current direction 
suggests that it was due to the magnetic inhomogeneities at the 
interfaces.  The hysteresis loops show that these inhomogeneities 
represent less than 2 \% of the total magnetization.  The amplitude of 
the observed effect and its dependence on the orientation of the wire 
in the magnetic field are in accordance with a model of magnetization 
reversal provoked by a transfer of moments from the spin polarized 
current into the magnetic wire.

\newpage 

\vspace{-0.7cm}

\vspace{-0.7cm}

\begin{thebibliography}{10}
\bibitem{sym} If the wires were perfect cylinders with 
identical ends, the symmetries of the system would impose that one 
has 
to reverse the sense of the current when reversing the field ramp in 
order to get the same effect.
\end{thebibliography}
\end{document}